\documentclass[twocolumn, prb,preprintnumbers,amsmath,amssymb,floatfix]{revtex4}
\usepackage[linktocpage,bookmarksopen,bookmarksnumbered]{hyperref}

\usepackage{graphicx}
\usepackage{dcolumn}

\usepackage{amsmath,graphics,epsfig,color,verbatim,ulem}
\usepackage{braket}
 
 \newcommand{\vR}{{\mathbf{R}}}
\renewcommand{\vr}{{\mathbf{r}}} 
\newcommand{\vk}{{\mathbf{k}}}
\newcommand{\vp}{{\mathbf{p}}}
\newcommand{\vG}{{\mathbf{G}}}
\newcommand{\vK}{{\mathbf{K}}} \newcommand{\vq}{{\mathbf{q}}}

\usepackage{tabularx,ragged2e,booktabs,caption}
\newcolumntype{C}[1]{>{\Centering}m{#1}}

\usepackage{multibib}

\begin{document}

\setlength{\pdfpageheight}{\paperheight}
\setlength{\pdfpagewidth}{\paperwidth}

\title{The Mott-semiconducting state in the magic angle bilayer graphene}
\author{M. Haule, E. Y. Andrei and K. Haule}
\affiliation{Department of Physics and Astronomy, Rutgers University, Piscataway, NJ 08854, United States.}
\begin{abstract}
Using non-perturbative theoretical method, we address the problem of strong correlations in twisted bilayer-layer graphene
at the magic angle. We concentrate on the solution without symmetry breaking, where conventional Mott insulating state is expected for all integer fillings. 
At Coulomb repulsion corresponding to dielectric constant $\varepsilon\approx 5$ and several integer fillings we find a Mott-semiconducting state, which simultaneously hosts the Mott state, and inside the Mott gap, a second much smaller semiconduting gap. The presence of these Mott-ingap states, which are located at the $\Gamma$ point, makes the Mott state strongly temperature dependent and leads to a bad-metal phase at elevated temperatures. 
The system is insulating at the charge neutrality point and at even fillings away from it. 
\end{abstract}
\date{\today}
\maketitle

%
%
%
%

Recent theoretical predictions of very narrow bands in twisted bilayer graphene (TBG) for certain ``magic angles''~\cite{PNAS_MacDonald,PhysRevB.82.121407,PhysRevB.93.235153,PhysRevB.85.195458} and the subsequent discovery of strong correlations~\cite{NatureTBG0,yankowitz2018tuning}, and superconductivity~\cite{NatureTBG} at the first magic angle, has spurred new interest in graphene and its superlattice. For the many-body community, this system represents a unique opportunity to better understand the effects of strong correlations and their dependence on the parameters in the theory, because TBG has several unique knobs to control the system, such as the bandwidth and doping, which both can be continuously changed with 
the electric field and the twist angle, without introducing disorder. Thus far  correlation effects in TBG have been addressed by the mean-field and Hartree-Fock methods~\cite{recent_MacDonald,triangularL,VishwanathSenthil}, however the non-perturbative methods, which are required to describe the Mott insulating phase, have not been applied to this system. Here we fill this void, and develop the concepts that allow us to employ the embedded Dynamical Mean Field Theory~\cite{Kotliar_rmp06,JPSJ} to this problem.

To describe correlations of an electronic system, it is essential to find a set of quantum wave functions in which electrons tend to slow down, and in which the Coulomb interaction is strong. Such a set of localized wave functions form a minimal basis to express the essential parts of the potential energy of the Hamiltonian.
%
On the other hand, the kinetic energy part, with its mean field potential, does not need to be given in the same basis, and is more efficiently expressed in a complete basis, such as the plane wave basis.
This flexibility is most commonly explored in solids with active $f$-orbitals (such as heavy fermions)~\cite{ourScience}, or in charge transfer systems with active $p$ and $d$ orbitals (such as the late transition metal oxides)~\cite{ZZA} in which a Hubbard-like model built from the very narrow $f$  or $d$ -orbitals does not describe the low energy physics well, while the generalized Anderson lattice model, or the $p$-$d$ model, are much more successful~\cite{Kotliar_rmp06}. The Coulomb interaction is strong when the electron resides on the $f$ (or $d$) orbital, which is localized in real space, but is much weaker for electrons in the itinerant $sp$ states, which are extended in space. 
When this flexibility is explored in TBG, one needs to find only a set of localized wave functions, centered on the $AA$ site of the Moir\'{e} lattice, which have large overlap with the low-energy narrow bands. Crucially, one does not need to faithfully describe the kinetic energy part of the Hamiltonian. Indeed, it was shown in Ref.\onlinecite{Vishwanath1,Shiang} that any set of atomic Wannier orbitals centered at the $AA$-site cannot describe the four low energy bands of the TBG, and in particular that a subset of the four bands at the $\Gamma$ point does not have overlap with any wave function centered at the AA site~\cite{Shiang,VishwanathSenthil,VafekWannier}. This is reminiscent of the $f$-systems, in which the itinerant states do not 
have a simple description in terms of localized wave functions. 

\begin{figure*}[bht]
\includegraphics[width=0.85\linewidth]{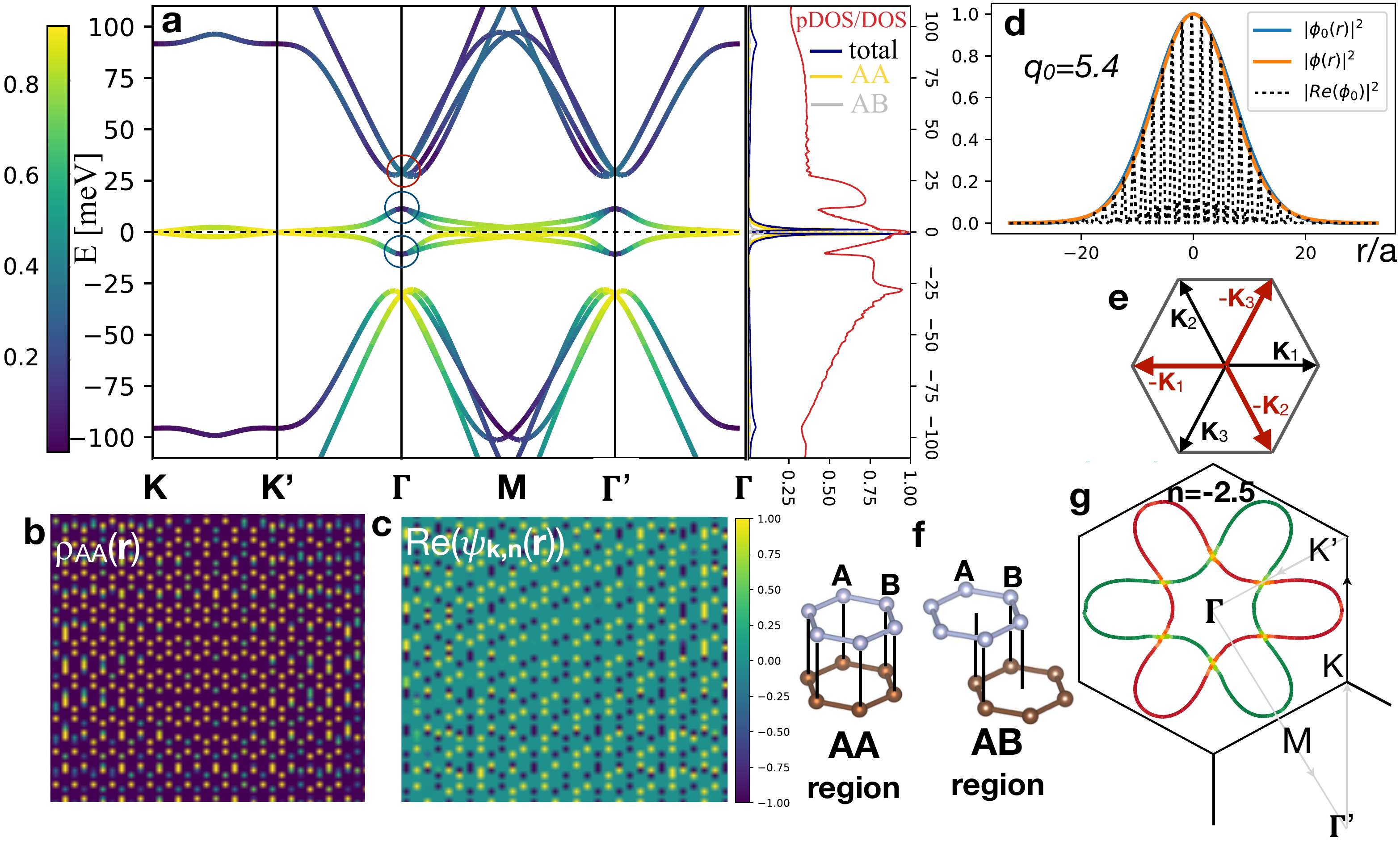}
\caption{ \textbf{The tight-binding band structure:} 
a) left: The dispersion of the low-energy bands. The color is proportional to the overlap of bands and the localized wave functions $\phi^{(1)}$..$\phi^{(4)}$. right: the corresponding total density of states (DOS) and the partial DOS (pDOS), and the ratio between pDOS and DOS. pDOS is obtained by projecting to functions $\phi^{(1)}$..$\phi^{(4)}$. The ratio pDOS/DOS is over 90\% for the low energy states b) the charge density in the real space close to the center of the $AA$ region,
c) the real part of the band eigenvector in the same real space, and at a generic k-point $\psi_{k,n}(\vr)$. It shows the phase, which corresponds to the momentum of the Dirac points of the single graphene layer. 
d) The real space gaussian localized wave function centered at the $AA$-site. 
e) The first Brillouin zone of the single-layer graphene, with the three $K$ vectors pointing towards the Dirac cones. 
f) The carbon configuration between the two layers at the center of the $AA$ and $AB$ regions.
g) The Moir\'{e}-Brillouin zone with the Fermi surface of the system at hole doping of 2.5 below the neutrallity point and the momentum-path used in a). The red (green) color corresponds to orbital 1 and 2 (3 and 4).
}
\label{fig1}
\end{figure*}
In the Moir\'{e} superstructure three regions can be identified, the $AA$ region where carbon $A$ atoms in both layers sit on top of each other (see Fig.~\ref{fig1}f), the $AB/BA$ regions, where the $A$ atoms in top layer are above the $B$ atoms of the bottom layer, and the bridge between the two.~\cite{PNAS_MacDonald} The low energy charge is very strongly concentrated on $AA$ sites, which form the triangular lattice.~\cite{PNAS_MacDonald} 
To describe the Coulomb interaction in this system we found four localized orbitals, which are centered at the $AA$ site, and contain most of the low energy electronic weight.
In Fig.~\ref{fig1}a we show the band structure where the color coding shows the amount of overlap between the bands and the localized wave functions. On the right, we display the total and the partial density of states (DOS), and their ratio. The partial DOS is obtained by projecting DOS to the localized functions $\phi_0^{(1)}\cdots \phi_0^{(4)}$ defined below. The plot  Fig.~\ref{fig1}a  shows that over 90\% of the low energy spectral weight is represented by these localized wave functions.
In Fig.~\ref{fig1}b we show the electronic charge as obtained from the tight-binding model in real space for the four low-energy bands near the center of the $AA$ region. It is centered on the triangular lattice~\cite{triangularL,NatureTBG} and quite strongly localized at the center of the $AA$ region (not shown). Moreover, the band eigenvectors ($\psi_{k n}(\vr)$ defined by $H\psi_{k n}(\vr)=\varepsilon_{k n}\psi_{k n}(\vr)$) have non-trivial phase, which is varying on the atomic scale distance (see Fig.~\ref{fig1}c), and needs to be properly accounted for when constructing localized wave functions. A close examination shows that the emergent periodicity corresponds to the wave vectors of the Dirac points of the single layer graphene, i.e., $\vK$ and $\vK'$~\cite{PNAS_MacDonald}. The four low-energy bands thus emerge from the standing waves between Dirac cones of the two graphene layers, as already discussed in Ref.\onlinecite{PNAS_MacDonald}, and can be represented by the following set of orthogonal functions, that are centered at the Moir\'{e} $AA$ site:
\begin{eqnarray}
\phi_0^{(1)}(\vr) &=& \phi_0(r) \frac{1}{\sqrt{3}} \sum_{p,j=1..3} e^{i\vK^p_j(\vr-\vR^p_A)} \nonumber\\
\phi_0^{(2)}(\vr) &=& \phi_0(r) \frac{1}{\sqrt{3}} \sum_{p,j=1..3} e^{i\vK^p_j(\vr-\vR^p_B)} \nonumber\\
\phi_0^{(3)}(\vr) &=& \phi_0(r) \frac{1}{\sqrt{3}} \sum_{p,j=1..3} e^{-i\vK^p_j(\vr-\vR^p_A)} \nonumber\\
\phi_0^{(4)}(\vr) &=& \phi_0(r) \frac{1}{\sqrt{3}} \sum_{p,j=1..3} e^{-i\vK^p_j(\vr-\vR^p_B)} \nonumber
\end{eqnarray}
where $\vK^p_j$ are the three vectors to the equivalent Dirac cones of the single layer graphene (see Fig.~\ref{fig1}e), $p$ stands for the top or bottom layer, and 
$\phi_0(r) \propto \exp(-(r \,q_0/R_{AA})^2)$ is the gaussian of width $R_{AA}/q_0$, and $R_{AA}$ is the separation between the $AA$ Moir\'{e} sites, and $q_0$ is a constant, which is optimized to achieve the best overlap with the low-energy band structure. $\vR^p_A$ and $\vR^p_B$ is the position of the two carbon atoms in the first graphene unit cell.
Note that $\phi_0^{(2)}$ and $\phi_0^{(4)}$ ($\phi_0^{(1)}$ and $\phi_0^{(3)}$) vanish on the $A$ ($B$) sub-lattice of a single layer graphene.

\begin{figure*}[bht]
\includegraphics[width=0.95\linewidth]{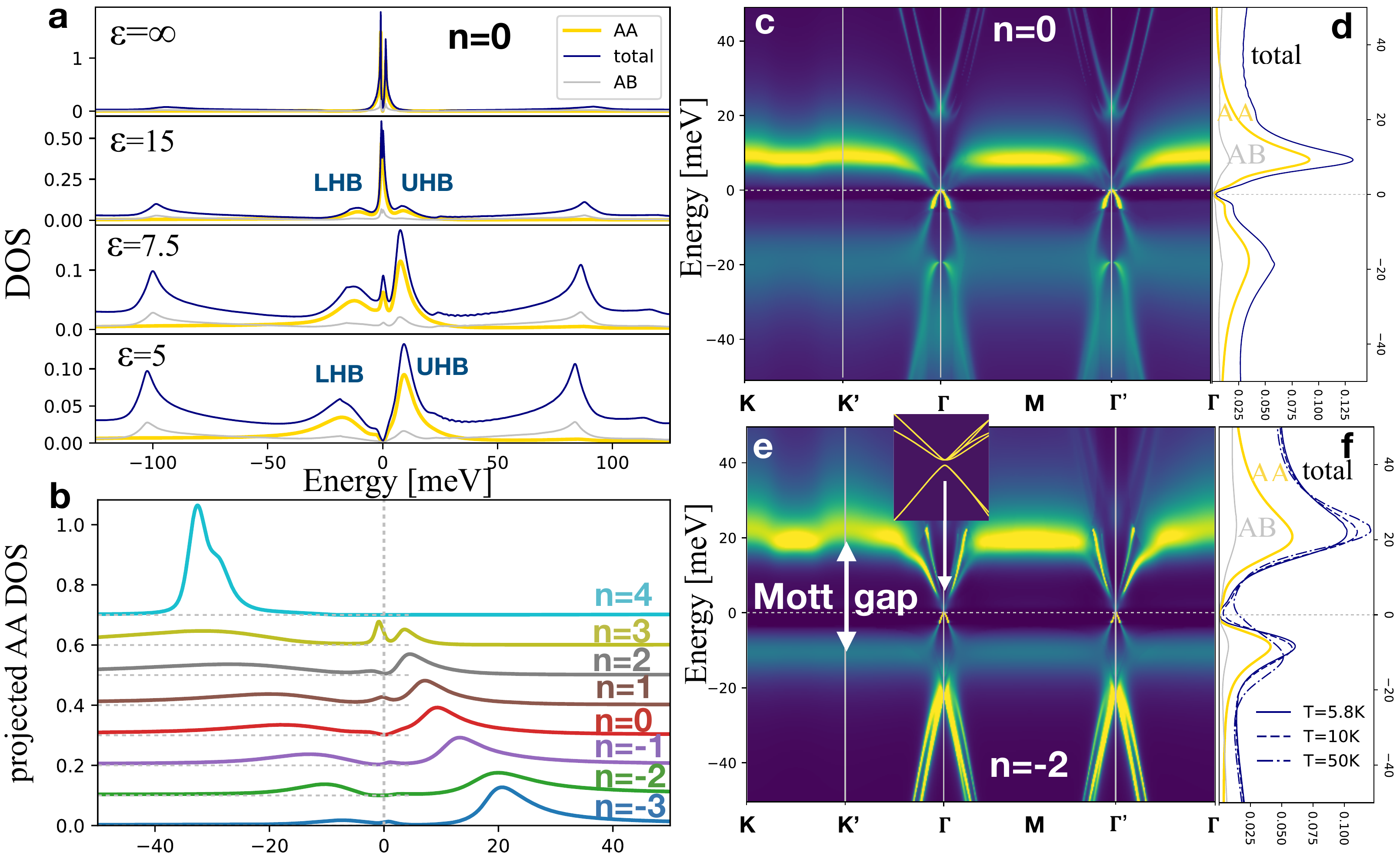}
\caption{ \textbf{DMFT spectra:}  The partial and the total DOS for various strength of the interactions at the half-filling $n=0$ ($\varepsilon=\infty$, $\varepsilon=15$, $\varepsilon=7.5$, $\varepsilon=5$ correspond to $U_0=0$, $U_0=50\,$meV,$U_0=100\,$meV and $U_0=150\,$meV, respectively ). The upper (UHB) and the lower Hubbard bands (LHB) are marked. b) Projected DOS for various integer fillings away from the charge neutrallity point at $U_0=150\,$meV. c) and e) momentum resolved spectra $A(\vk,\omega)$ at charge neutrallity point ($n=0$) and at hole doping $n=-2e$ ($U_0=150\,$meV. d) and f) The corresponding partial and total DOS.
}
\label{fig2}
\end{figure*}
We construct the final DMFT projector wave-functions $\phi^{(i)}(\vr)$ by projective-orthogonalization, i.e., 
\begin{eqnarray}
\ket{\phi^{(i)}} =  \sum_{n\in w} \ket{\psi_{kn}}\braket{\psi_{kn}|\phi_0^{(j)}}\left(\frac{1}{\sqrt{O}}\right)_{ji}
\end{eqnarray}
in which the sum over bands is extended to a large energy window $w$ (here 400$\,$meV), spanning substantially larger energy window than the value of the Coulomb repulsion (see supplementary material~\cite{suppl}). $O$ is the overlap $\sum_n\braket{\phi_0|\psi_{k n}}\braket{\psi_{k n}|\phi_0}$. When the parameter of the real-space extent ($q_0$) is optimized so that the localized wave functions represent well the electronic charge in the $AA$ region, the difference between the trial function $\phi_0^{(j)}(\vr)$ and the final projective $\phi^{(j)}(\vr)$ is minimal. Fig.~\ref{fig1}d shows that the two functions are almost indistinguishable, which can be achieved with the large energy windows $w$, such that the orthogonalization does not substantially alter the shape of the localized wave function. This is important because it facilitates computing the matrix elements of the Coulomb repulsion. 

To estimate if an electron can get localized on a given orbital and site, it is sufficient to look at its DMFT-hybridization function~\cite{suppl},  which describes how easy it is for an electron to espace from  this orbital and this site. If the hybridization function is small, and the density of states is large, a modest Coulomb repulsion can give divergent correlations, and consequently Mott localization. This happens on the $AA$-sites when the dielectric constant is around $\varepsilon\approx 5$ (see below). We also construct similarly localized wave functions on the $AB$ and $BA$ sites, but we find that 
the same amount of the Coulomb repulsion does not affect the electrons on the $AB$ and $BA$ sites, because their hybridization function is approximately ten-times larger than on the $AA$-site (see supplementary~\cite{suppl}).
We note in passing that these four localized wave functions (or eight if $AB$, $BA$ functions are included) are not sufficient to faithfully represent the kinetic part of the Hamiltonian, hence they do not constitute a complete set of Wannier functions. To complete the basis, one would need to include several additional Wannier functions centered on other sites in real space~\cite{Shiang}. However, these other Wannier functions are much less localized, and consequently the effect of the Coulomb repulsion is small, and will be neglected here.

Within Dynamical Mean Field theory, the kinetic energy part of the Hamiltonian can be evaluated in the complete tight-binding plane wave basis, while the dynamic effects of the Coulomb repulsion are considered when electrons sit on the localized $\ket{\phi^{(j)}}$ wave functions. 
The matrix elements of the Coulomb repulsion between $\ket{\phi_0^{(j)}}$ functions can be readily evaluated, and given that $\ket{\phi^{(j)}}$ are very similar to $\ket{\phi_0^{(j)}}$, it is an excellent approximation to use these matrix elements to
construct the potential part of the Hamiltonian. First we evaluate the direct terms of the Coulomb repulsion, 
$U_0\equiv U_{ijji}=\braket{\phi_0^{(i)}\phi_0^{(j)}|\frac{e^2}{\varepsilon|\vr-\vr'|}|\phi_0^{(j)}\phi_0^{(i)}}=\frac{e^2 \sqrt{\pi}(\sqrt{2}-1)q_0}{\varepsilon R_{AA}}\approx \frac{136.6\,\textrm{meV} q_0}{\varepsilon}$. Given that optimal $q_0\approx 5.4$ and dielectric constant $\varepsilon$ of graphene is typically estimated to be around $5$~\cite{recent_MacDonald}, the reasonable value of the Coulomb repulsion is of the order of $150\,$meV. The biggest surprise is that the Hund's interaction terms, while nonzero, are extremely small in this system. In particular $U_{1212}=U_{3434}=0$ because one of the two functions is finite only on the $A$ sites and the other on the $B$ sites. The values of $U_{1414}=U_{2323} \approx 0.023 U_0$ is also very small because different functions have different phase factors, hence the Hund's interaction is reduced approximately by the factor $\frac{1}{N_{cell}}\sum_{\vR_A\in Moire-cell}e^{i(\vK-\vK')\vR_A}$, making it substantially smaller than in typical solid state systems. Indeed the ratio between the Hund's term and the Hubbard term in most solids ranges between $1/3-1/10$, while here it is $U_{1414}/U_0 \approx 1/43$, hence this system is extremely close to the degenerate form of the Coulomb interaction, with an overwhelmingly Hubbard-type interaction.

The Hamiltonian (see supplementary~\cite{suppl}) defined by the kinetic term from the tight-binding approximation~\cite{PRB_MacDonald,PNAS_MacDonald}, and the Coulomb interaction on the localized $AA$ sites, can now be solved by the Dynamical Mean Field Theory~\cite{Kotliar_rmp06,JPSJ}, i.e., treating the local correlations exactly. In this work we concentrate on the paramagnetic solutions in the absence of long-range order symmetry breaking, although it is likely that at low temperature magnetic long range order or other type of charge or orbital order can be stabilized~\cite{triangularL}. To account for the lattice relaxation, we followed Ref.\onlinecite{Vishwanath} and reduce the interlayer tuneling amplitude between $A$-$A$ sites as compared to $A$-$B$ sites such that $w_{A-A}/w_{A-B}=0.75$ (see~\onlinecite{suppl}). As in Refs.~\onlinecite{Vishwanath,PNAS_MacDonald} the interlayer hopping is set to $w_{A-B}=110\,$meV. The temperature is set to $5.8\,$K, unless stated otherwise.

Fig.~\ref{fig2}a shows the local spectral function for different values of $U_0$ from 0-150$\,$meV at magic angle $\theta=2\,\arctan(1/( 63\sqrt{3})) \approx 1.05^\circ$ and at the charge-neutrallity point, which we will denote by $n=0$. Positive $n$ corresponds to electron and negative to hole doping away from the half-filled system. Without considering the Coulomb repulsion, the splitting of the two peaks and the half-bandwidth of a single peak is extremely small ($2.2\,$meV and $1.5\,$meV) and the bands touch at $A$ and $B$ points in the Moir\'{e} zone (see Fig.~\ref{fig1}). A small Coulomb repulsion of $50\,$meV (corresponding to the dielectric constant $\varepsilon=15$) first creates a three peak structure with Hubbard bands roughly $10\,$meV away from the central peak, and a double-split central peak, which gives rise to a bad metallic behaviour. It has vanishing density of states only at zero temperature, but due to finite electron-electron scattering rate, there is quite large number of states at the Fermi level at finite temperature. Increasing the interaction ($U=100\,$meV) reduces the strength of the central metallic peak to benefit the Hubbard bands and because the scattering rate increases, the splitting of the quasiparticle peak can no longer be resolved at this temperature. Finally, at Coulomb strength of $U_0=150\,$meV, corresponding to dielectric constant of $\varepsilon=5$, the central peak completely disappears and the Hubbard bands splitting and width increases to approximately $35\,$meV and $20\,$meV. We note that these numbers sensitively depend on the value of the tight-binding tunneling matrix elements, and even a small change in their values can substantially increase this width, as shown in Ref.\onlinecite{STM_exp}, hence these values should give only a correct order of magnitude, while their values will change once more realistic tight-binding model, which accounts for the lattice relaxation, is developed~\cite{PhysRevB.98.235404}. 

In Fig.~\ref{fig2}b we show the DMFT solution for other integer fillings away from the charge neutrallity point. With hole doping, the upper Hubbard band gains extra weight, and consequently the lower must loose some weight. The dopings of $+1$ and $-1$ and $+3$ and $-3$ away from the charge neutrality point are not truly insulating, although at all integer dopings the density of states is very small, as compared to doped system (see supplementary~\cite{suppl}). Most interesting state is found at $-2$, $0$, and $+2$ doping, which have a true gap, and the Mott gap is strongest at $-2$ doping.
The most surprising result is that this state is not a regular Mott insulator, as it is clear from approximately V-shaped DOS at the Fermi level. 
The momentum resolved spectral function in Fig.~\ref{fig2}c and e shows that while most of the spectral weight is pushed away from the Fermi level by strong interactions, there is a set of low energy bands left inside the Mott gap.
The part below the Fermi level has a very sharp bright part, which is cut-off at approximately $4\,$meV by the pole in the self-energy. Namely, the Mott-insulating state is characterized by the electronic self-energy which has a singular pole inside the Mott gap, and its imaginary part is very small in the rest of the gap. Hence, the electronic state is very sharp when the scattering rate is small, but is cut-off at the energy of the self-energy pole. The states near and above the Fermi level are less well defined and are also mostly residing away from the $AA$ region in real space.
The low-energy states have a particular topology shown in the inset of Fig.~\ref{fig2}e. 
These bands are obtained by setting the self-energy $\Sigma(\omega)\rightarrow \Sigma(0)$. The valence band is doubly degenerate at the $C=\Gamma$ point, and the degeneracy is protected by $C_3$ symmetry. The states mostly come from the bands encircled by blue oval in Fig.~\ref{fig1}a. On the other hand, the conduction bands have degeneracy four at $\Gamma$ point, and their degeneracy is protected by different symmetry, namely the time reversal and C$_2$. For example, if we make all four self-energies, that correspond to $\phi^{(1)}..\phi^{(4)}$ to be different, all degeneracies in the conduction bands are lifted, while the valence bands remain degenerate. The time reversal symmetry breaking splits four-time degeneracy into two doubly-degenerate sets. The states in the conduction band come primarily from bands encircled by red oval in Fig.~\ref{fig1}a.

We notice in passing that all those bands left over inside the Mott gap have very small overlap with the states at the $AA$ site, hence they come from more itinerant states from different parts of the real space. The low energy semiconducting gap size is only $\approx 2$meV, much smaller than the 
Mott  gap ( $\approx 30\,$meV).
This unusual Mott electronic state has some similarity with the orbitally selective Mott state, in which some orbitals are Mott insulating and others are not, but it is different from it, because the states in the gap do not cross the Fermi level, and do not have a finite Fermi surface, hence the system is still insulating. However, the dispersive states do play an important role at the elevated temperature. Once the scattering rate at zero frequency becomes comparable to the small gap, the hybridization function becomes finite, and one starts to see metallic states in the gap. This is similar to behaviour of correlated semiconductors with a small gap, such as FeSi~\cite{FeSius}, where the semiconducting gap is filled-in when temperature is increased. We show the calculated density of states for three temperatures ($5.8\,$K, $10\,$K and $50\,$K) in Fig.~\ref{fig2}f. The Mott gap does not collapse, but it is filled-in by the incoherent weight, which allows one to move the chemical potential into the itinerant states, making the system effectively metallic. This metallization with increasing temperature might be able to explain a surprising finding in Ref.~\onlinecite{NatureTBG0} that the conductance changes slope with increasing temperature, making the system metallic at elevated temperatures.

We acknowledge the support of NSF DMR-1709229 (K.H.) and NSF DMR 1708158 (E.Y.A)

\bibliography{reftbg}
\bibliographystyle{apsrev4-1}
\newpage
\section*{Supplementary}

\subsection{Kinetic part of the Hamiltonian}
To construct the tight-binding model, we followed Refs.~\cite{PRB_MacDonald,PNAS_MacDonald} and approximate the Fourier transform of the interlayer tunneling by 
\begin{eqnarray}
t(q) = t_0 e^{-\alpha (q d)^\gamma}
\end{eqnarray}
with $t_0=1.066\,$eV, $\alpha=0.13$, $\gamma=1.25$, and the interlayer distance $d=3.34\textrm{\AA}$, so that the important parameter, which determines the low energy bandwith $w_{A-B}\equiv t(k_D=\frac{4\pi}{3 a})=110\,$meV (see Refs.~\cite{PNAS_MacDonald,Vishwanath}), and $a=1.42\textrm{\AA}$ is the graphene lattice spacing. The hopping within the graphene layer is set to $t_{gr}=2.73\,$eV. 

As derived in Ref.~\onlinecite{PNAS_MacDonald}, the tunneling matrix elements can be expressed by
\begin{widetext}
\begin{eqnarray}
T_{\vk,\vp}^{\alpha,\beta}=\braket{\psi^t_{\vk,\alpha}|H_{tun}|\psi^b_{\vp,\beta}}=
\sum_{\vG^t,\vG^b} t(|\vk+\vG^t|)\delta(\vp+\vG^b-\vk-\vG^t) e^{i\tau_\alpha \vG^t-i\tau_\beta\vG^b}
\label{on1}
\end{eqnarray}
\end{widetext}
where $\vG^s$ are reciprocal vectors of a single layer graphene, and $s=[t,b]$ marks the top or bottom layer. $\vk$, $\vp$ are momentums in the Bouillon zone of the top and the bottom graphene layer, respectively. $\tau_\alpha$ is the position of the carbon atom within the unit cell, and $\alpha,\beta$ can be either $A$ or $B$ carbon atoms.
This is straightforwardly derived from the tight-binding solutions of a single graphene layer:
\begin{eqnarray}
\ket{\psi^{s}_{\vk}}=\frac{1}{\sqrt{N}}\sum_{\vR^s\in triang} e^{i\vk(\vR^s+\tau^s_\alpha)}\ket{\vR^{s}+\tau_\alpha^s}
\end{eqnarray}
and from the form of the tunneling matrix elements, which depend only on the distance between atoms, i.e.,
\begin{widetext}
\begin{eqnarray}
\braket{\vR^t+\tau_\alpha^t|H_{tun}|\vR^b+\tau_\beta^b}=t_r(|\vR^t+\tau_\alpha^t-\vR^b-\tau_\beta^b|)=\sum_{\vq^t,\vG^t} t(|\vq^t+\vG^t|) e^{i(\vq^t+\vG^t)(\vR^t+\tau_\alpha^t-\vR^b-\tau_\beta^b)}
\end{eqnarray}
\end{widetext}
The last identity is just the exact Fourier transform of the real-space hopping matrix elements $t_r(r)$.

In Eq.~\ref{on1} each contribution to the hopping $|T^{A,A}_{\vk,\vp}|$ is equal to $|T^{A,B}_{\vk,\vp}|$, which is an artifact of negligence of lattice relaxation~\cite{Vishwanath}. Once the lattice is allowed to relax, each contribution to $|T^{A,A}|$ and $|T^{A,B}|$ differ, and as suggested by 
Ref.~\onlinecite{Vishwanath} we reduced  diagonal contributions by 25\%, i.e.,
\begin{eqnarray}
T_{\vk,\vp}=\sum_{\vG^t,\vG^b} 
e^{-\alpha (|\vk+\vG^t| d)^\gamma}\delta(\vp+\vG^b-\vk-\vG^t) 
\nonumber\\
\left(
\begin{array}{cc}
0.75\,t_0\, e^{i\tau_A(\vG^t-\vG^b)} & t_0\, e^{i\tau_A \vG^t-i\tau_B\vG^b}\\
t_0\, e^{i\tau_B \vG^t-i\tau_A\vG^b} & 0.75\,t_0\, e^{i\tau_B (\vG^t-\vG^b)} 
\end{array}\right),
\end{eqnarray}
which opens a decent band gap between the four low energy bands and the rest of the higher-energy bands.

Now that we have the tunneling matrix elements, we need to diagonalize a large Hamiltonian matrix. To do that,
we notice that all momentum vectors can be expressed in terms of integer multiples of the Moir\'{e}-Brillouin zone (MBZ) reciprocal basis, plus a vector inside the first MBZ.
Since MBZ is the emergent Brillouin-zone, tunneling can not mix different momentum vectors of the MBZ. Therefore both $\vk$  and $\vp$ in $T_{\vk,\vp}$ correspond to the same MBZ momentum, and can be expressed
\begin{eqnarray}
\vk = \vk_{MBZ} + \vec{b}_1\; n_k^1  + \vec{b}_2\; n_k^2\\ 
\vp = \vk_{MBZ} + \vec{b}_1\; n_p^1  + \vec{b}_2\; n_p^2
\end{eqnarray}
where $n_k^i$ are integers and $\vec{b}_i$ are reciprocal basis vectors of MBZ. Furthermore, it is easy to see that any commensurate angle requires that the reciprocal vectors of the top and the bottom layer are related by integers to the MBZ reciprocal vectors $\vec{b}_i$, i.e.,
\begin{eqnarray}
\left(
\begin{array}{c}
\vec{b}_1^t\\
\vec{b}_2^t
\end{array}
\right)=
\left(
\begin{array}{cc}
n^t_{11} & n^t_{12}\\
n^t_{21} & n^t_{22}
\end{array}
\right)
\left(
\begin{array}{c}
\vec{b}_1\\
\vec{b}_2
\end{array}
\right)
\end{eqnarray}
and similarly for the bottom layer.
The tunneling matrix elements $T_{\vk,\vp}$ require one to find \textit{all vectors}, which satisfy $\vp+\vG^b-\vk-\vG^t=0$.
This can be expressed as
\begin{eqnarray}
\vp -\vk + \vec{b}_1^b\; m_1^b+ \vec{b}_2^b\; m_2^b - \vec{b}_1^t\; m_1^t- \vec{b}_2^t\; m_2^t =0
\end{eqnarray}
Because $\vp$ and $\vk$ share the same $\vk_{MBZ}$ and because all other momenta are integer multiple of $\vec{b}_i$, we see that the condition to find integers $m_i^s$ is a special case of so-called system of linear diophantine equations. First, let us write the two systems of equations
\begin{widetext}
\begin{eqnarray}
(n_p^1-n_k^1,n_p^2-n_k^2)
\left(
\begin{array}{c}
\vec{b}_1\\
\vec{b}_2
\end{array}
\right)
+ 
(m_1^b, m_2^b)\cdot
\left(
\begin{array}{cc}
n^b_{11} & n^b_{12}\\
n^b_{21} & n^b_{22}
\end{array}
\right)
\left(
\begin{array}{c}
\vec{b}_1\\
\vec{b}_2
\end{array}
\right)
-
(m_1^t, m_2^t)\cdot
\left(
\begin{array}{cc}
n^t_{11} & n^t_{12}\\
n^t_{21} & n^t_{22}
\end{array}
\right)
\left(
\begin{array}{c}
\vec{b}_1\\
\vec{b}_2
\end{array}
\right)=0
\end{eqnarray}
or
\begin{eqnarray}
(m_1^b, m_2^b, -m_1^t, -m_2^t)\cdot
\left(
\begin{array}{cc}
n^b_{11} & n^b_{12}\\
n^b_{21} & n^b_{22}\\
n^t_{11} & n^t_{12}\\
n^t_{21} & n^t_{22}
\end{array}
\right)=(n_k^1 -n_p^1, n_k^2 -n_p^2) 
\end{eqnarray}
\end{widetext}
This defines the system of linear diophantine equations ($\vec{x}\cdot A=\vec{b}$ or $A^T\cdot \vec{x}=\vec{b}$), which can be solved by transforming the $4\times 2$ matrix on the left into the Hermite-Normal form. The corresponding pivotal matrix then contains the two dimensional space of solutions to the homogeneous part of the equation. Thus, instead of searching for four dimensional vector $(m_1^b, m_2^b, -m_1^t, -m_2^t)$ through expensive looping over all integers, we can determine all possible solution at once by just precomputing the Hermite-Normal form of the matrix, which depends only on the tilt angle, but not on the momentum. For example, at the tilt angle $1.05^\circ$ the above matrix takes the form
\begin{eqnarray}
\left(
\begin{array}{cc}
  63 & 32\\
-32 & 31 \\
  63 & 31 \\
-31 & 32
\end{array}
\right).
\end{eqnarray}

Finally, the number of possible momenta $\vp$ and $\vk$ which can couple by tunneling is finite, as is given by the tilt angle. For example, at angle $1.05^\circ$ there is only $2977$ $\vk$ vectors and $2977$ $\vp$ vectors, which can couple, hence the resulting matrix is of size 11908. This can be straightforwardly diagonalized by LAPACK libraries. However, we found that the matrix size can be considerably reduced without loss of accuracy for the energy range we are interested in, by removing parts of the Hamiltonian matrix, which have very large diagonal energy.

\subsection{Total Hamiltonian}

\begin{figure}[bht]
\includegraphics[width=0.9\linewidth]{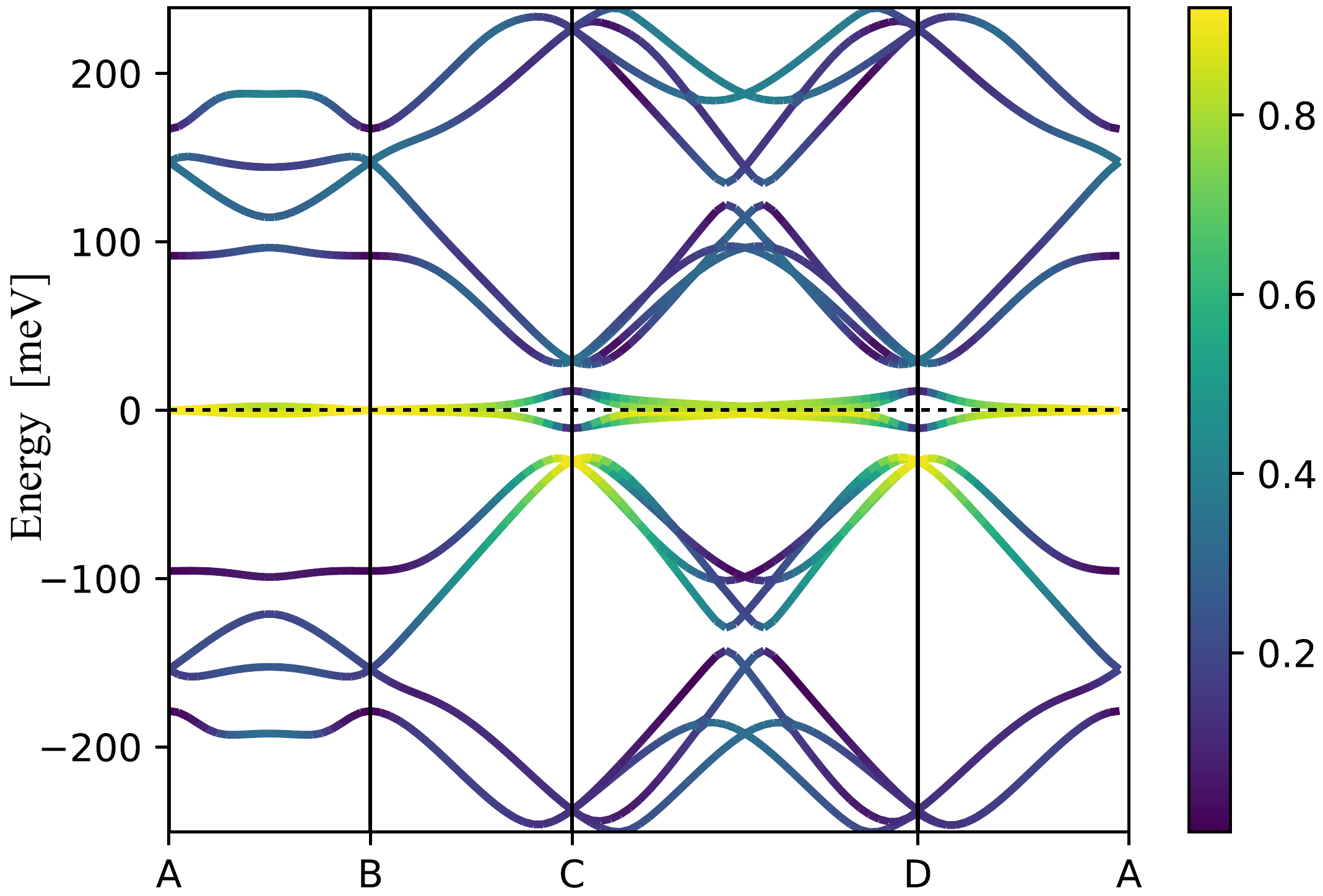}
\caption{ \textbf{The band structure: }  in the extended energy window, in which $\phi^{(j)}$ functions are orthogonalized.
The yellow color shows where the $\phi^{(j)}$ functions have large overlap with the bands.
}
\label{fig3}
\end{figure}

Once the tight-binding matrix (at $1.05^\circ$ of size $11908\times 11908$) is diagonalized, we obtain a set of bands $\varepsilon_{\vk,n}$ and the corresponding eigenvectors 
$\psi_{\vk n s}$, which define the kinetic part of the Hamiltonian. Then we compute the projection of the localized $AA$-centered functions to the band eigenvectors $\braket{\psi_{kn}|\phi^{(j)}}$, and we orthogonalized $\phi^{(j)}$ functions in an extended energy window. In this work, we use 20 bands, which span over $400\,$meV energy window, depicted in Fig.~\ref{fig3}.
To carry out the orthonormalization given by Eq.1 in main text, we performed the singular value decomposition of the overlap, i.e, $\braket{\psi_{kn}|\phi_0^{(j)}}=U s V^+$. At the optimized parameter of the extent $q_0$, the 
smallest singular value $s$ reaches its maximum. Here overlap $\frac{1}{\sqrt{O}}=V \frac{1}{s} V^+$, and $U$, $V$ are unitary matrices. For example, at angle $1.05^\circ$, and window $w=400\,$meV, singular values $s$ range from 0.584  to 0.725 at $q_0=5.4$, indicating a good-quality choice for the localized wave functions.

\begin{figure}[bht]
\includegraphics[width=0.9\linewidth]{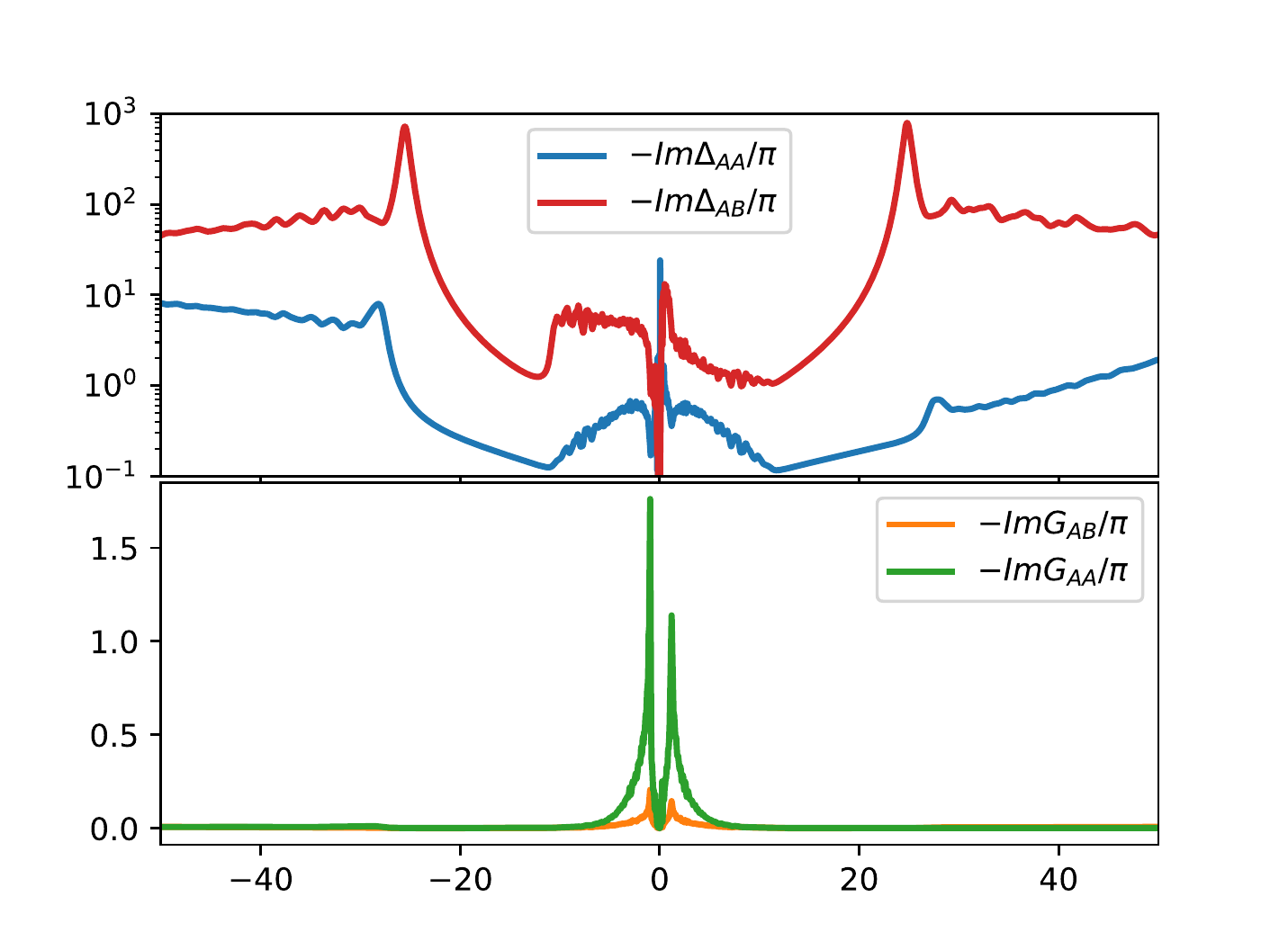}
\caption{ \textbf{The tight-binding $U=0$ hybridization function} in the low energy region, computed by 
$d(\omega)=-\frac{1}{\pi}Im\Delta(\omega)=\frac{1}{\pi}Im G^{-1}(\omega)$.
The top row shows the imaginary part of the hybridization function on the $AA$ and $AB$ sites. At low energy the $AA$ hybridization is of the order of $1\,$meV while $AB$ is of the order of $10\,$meV, which makes an enormous difference in the strength of correlations on the two sites.
The bottom shows the density of states on $AA$ and $AB$ sites 
}
\label{fig4}
\end{figure}

\begin{figure}[bht]
\includegraphics[width=0.9\linewidth]{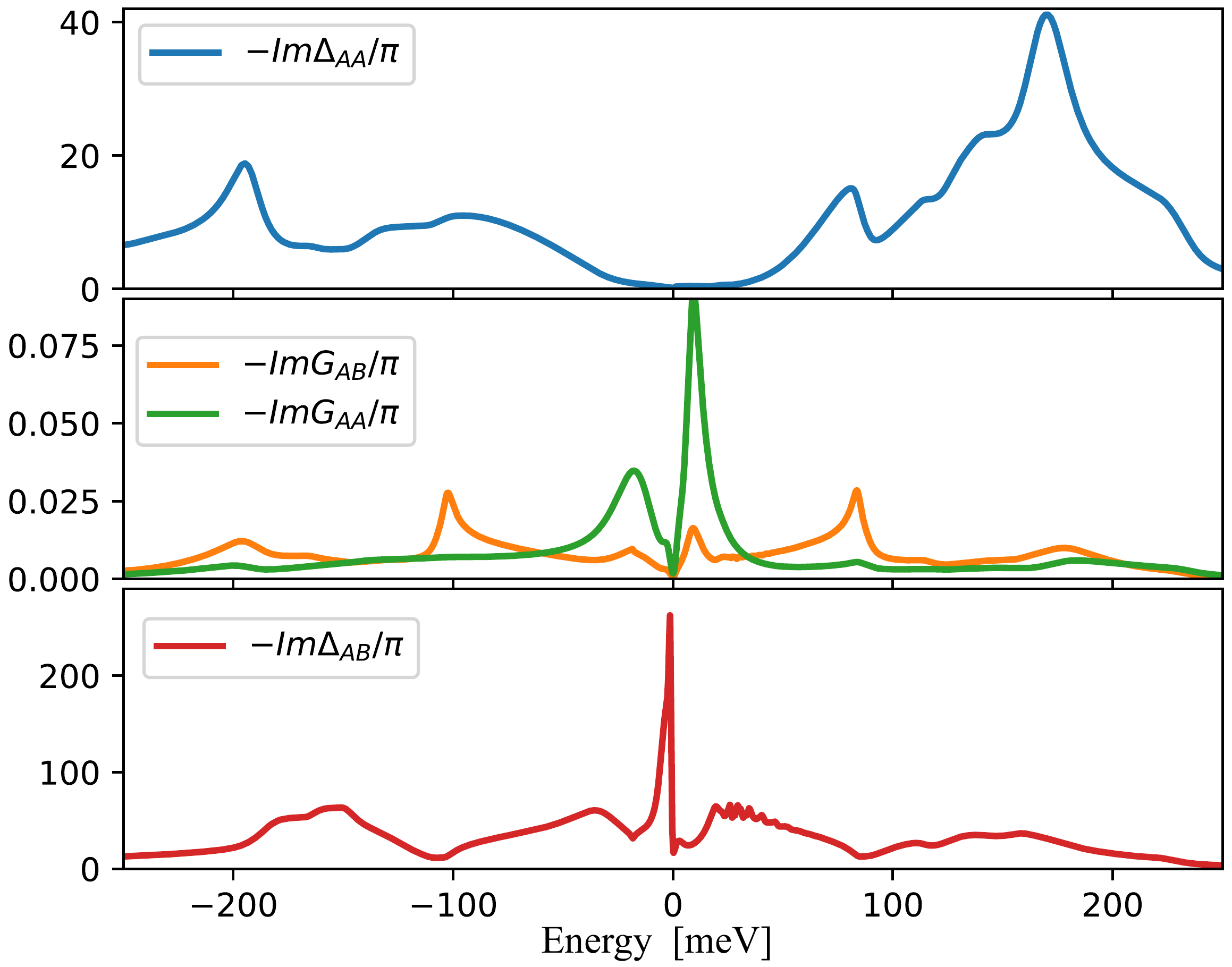}
\caption{ \textbf{The hybridization function} at $U_0=150\,$meV in an extended energy region. The top row shows the imaginary part of the hybridization function on the $AA$ sites. Noteworthy is its even smaller value than at $U_0=0$ due to opening of the correlation gap.
The middle shows the density of states on $AA$ and $AB$ sites (at $n=0$). Clearly the $AA$ spectra is centered at low frequency, but has large tails. The $AB$ spectra is mostly concentrated in the site-peaks at higher energies.
Bottom: The hybridization on the $AB$ sites, which remain large even in the Mott states (note the difference in scale).
}
\label{fig5}
\end{figure}

The potential energy part contains the Coulomb interaction, which is written in the localized basis, centered on the $AA$, $AB$, and $BA$ sites. We found that Coulomb interaction on $AB$ and $BA$ sites is irrelevant, because the hybridization function is more than one order of magnitude larger than on the $AA$ site (see Figs.~\ref{fig4} and \ref{fig5}). Since the correlations strength is exponentially sensitive to the hybridization strength, the Coulomb repulsion, which localizes electrons on the $AA$-sites has negligible effect on electrons at the $AB$ and $BA$ sites.

Concentrating then on the interaction at the $AA$ site, the Hamiltonian takes the form:
\begin{widetext}
\begin{eqnarray}
H = \sum_{\vk,n,s} \varepsilon_{\vk n}\psi^\dagger_{\vk n s}\psi_{\vk n s} 
+ \sum_{\vR_{AA},s,s'}\sum_{i,j=1}^4 U_{ijji} \psi^\dagger_{\vR_{AA},i,s}\psi^\dagger_{\vR_{AA},j,s'}\psi_{\vR_{AA},j,s'}\psi_{\vR_{AA},i,s}\\
+ \sum_{\vR_{AA},s,s'}\sum_{i,j=1}^4 U_{ijij} \psi^\dagger_{\vR_{AA},i,s}\psi^\dagger_{\vR_{AA},j,s'}\psi_{\vR_{AA},i,s'}\psi_{\vR_{AA},j,s}
-U_0 \left(\sum_{\vR_{AA},i,s} \psi^\dagger_{\vR_{AA},i,s}\psi_{\vR_{AA},i,s}-\frac{1}{2}\right)
\end{eqnarray}
\end{widetext}
where $n$, $s$ denote the band and spin, $R_{AA}$ denotes the $AA$ sites, and $i,j$ run over the four correlated orbitals. Note that $\psi^\dagger_{\vR_{AA},i,s}$ creates an electron in the $\Phi^{(i)}(\vr-\vR_{AA})$ orbital, and $\psi^\dagger_{\vk n s}$ creates an electron in the tight binding band.
The last term is the double-counting term in the so-called localized limit. We checked that subtracting instead the Hartree-Fock energy does not change results appreciatively. The need to subtract the double-counting is in the fact that we are dealing with a beyond-Hubbard model system, where the double-counting can not be simply hidden into the shift of the chemical potential, and a proper alignment of the itinerant states with respect to the correlated orbitals can only be achieved realizing that the tight-binding model already contains the Hartree term, and the semilocal-part of the exchange-correlation energy. 
Note that the occupation of the local orbitals is not simply related to the occupation of the entire system, and that the gap opening occurs at integer filling of the total system, which does not coincide with the integer filling of the $AA$ local orbitals.
For example, at the charge neutrallity point $n=0$ the occupation of the local orbital is close to $n_{imp}=4$, as expected, however, at $n=-1$, $n=-2$ and $n=-3$ the occupation of the four localized orbitals is $n_{imp}=3.07$, $n_{imp}=2.13$, and  $n_{imp}=1.34$, respectively.

Finally, let us comment on the asymmetry of the density of states. The current tight-binding model is almost particle-hole symmetric at $n=0$ with slightly larger peak below $EF$ as compared to the one above $EF$ (see Fig.~\ref{fig2}a of the main text). Once the Coulomb interaction is turned on, the asymmetry is flipped, so that the peak below $EF$ is somewhat broader and lower. This is mainly because of the asymmetry in the hybridization function, with much larger peak of hybridization above $EF$  as compared to the peak below $EF$ (top panel in Fig.~\ref{fig5}). We note that this asymmetry is hard to guess from looking at the band structure or density of states. We also notice that this asymmetry will depend on the tight-binding model, and more realistic tight-binding model might change this asymmetry.

\begin{figure}[bht]
\includegraphics[width=0.9\linewidth]{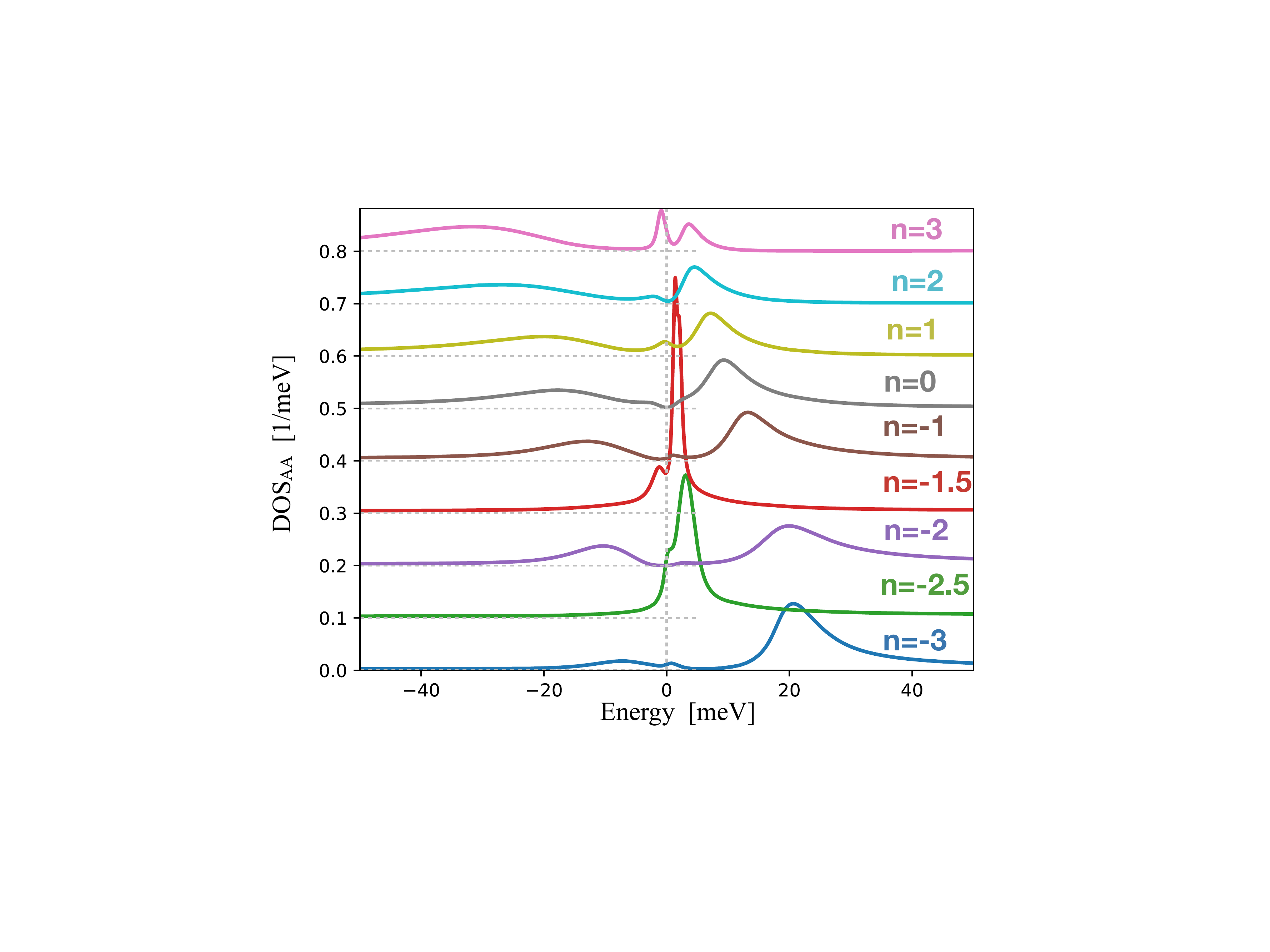}
\caption{ \textbf{DMFT density of states:}  projected to the four orbitals on the $AA$-site, for various integer and half-integer fillings. The integer fillings have either vanishing ($n=$0,$\pm$2,) or extremely small DOS ($n=\pm$1,$\pm$3), while half-integer fillings display much larger DOS at the Fermi level.
}
\label{fig6}
\end{figure}
In the main text we show DOS for integer fillings, but not for doping away from the integer fillings. In Fig.~\ref{fig6} we also show a few half-integer fillings, which display a very clear quasiparticle peak at the Fermi level. However, all quasiparticle peaks inherit a dip close to $EF$, which is inherited from the tight-binding model, in which the central peak is split into two peaks.
Finally, the system is pretty good fermi liquid at $n=-2.5$, hence the Fermi surface at this doping is very similar to the one given in Fig.1f of the main text with two overlaping fidget-spinners.
\end{document}